\documentclass[referee]{raa}            

\usepackage{graphicx,times}             
\usepackage{natbib}
\usepackage{amsmath}	
\usepackage{amssymb}	
\usepackage{xspace}
\usepackage{bigints}
\usepackage{siunitx}
\usepackage[figuresright]{rotating}
\usepackage{pdflscape}

\begin{document}

\title{Jets, accretion and spin in supermassive black holes}

\volnopage{ {\bf 2018} Vol.\ {\bf X} No. {\bf XX}, 000--000}
   \setcounter{page}{1}

\author{Yongyun Chen\inst{1,3}, Qiusheng Gu\inst{2,3}, Jianghe Yang\inst{4}, Junhui Fan\inst{5}, Xiaoling Yu\inst{1}, Dingrong Xiong \inst{6}, Nan Ding\inst{7}, Xiaotong Guo\inst{8}}

   \institute{College of Physics and Electronic Engineering, Qujing Normal
   	University, Qujing 655011, P.R. China: ynkmcyy@yeah,net \\
   	\and 
    School of Astronomy and Space Science, Nanjing University, Nanjing 210093, P. R. China:qsgu@nju.edu.cn\\
    \and
     Collaborative Innovation Center of Modern Astronomy and Space Exploration, Nanjing 210093, P. R. China \\   
     \and 
     College of Mathematics and Physics Science, Hunan University of Arts and Science, Changde 415000, People's Republic of China\\
     \and
     Center for Astrophysics,Guang zhou University,Guang zhou510006, China\\
     \and
     Yunnan Observatories, Chinese Academy of Sciences, Kunming 650011,China\\
     \and
     School of Physical Science and Technology, Kunming University 650214, P. R. China\\
     \and
     Institute of Astronomy and Astrophysics, Anqing Normal University, Anqing, Anhui 246133, People's Republic of China\\
{\small Received ********; accepted ********}  }

\abstract{The theoretical model suggests that relativistic jets of AGN rely on the black hole spin and/or accretion. We study the relationship between jet, accretion, and spin using supermassive black hole samples with reliable spin of black holes.
Our results are as follows: (1) There is a weak correlation between radio luminosity and the spin of black hole for our sample, which may imply that the jet of the supermassive black hole in our sample depends on the other physical parameters besides black hole spins, such as accretion disk luminosity. (2) The jet power of a supermassive black hole can be explained by the hybrid model with magnetic field of corona. (3) There is a significant correlation between radio-loudness and black hole spin for our sample. These sources with high radio-loudness tend to have high black hole spins. These results provide observational evidence that the black hole spin may explain the  bimodal phenomena of radio-loud and radio-quiet AGN.
\keywords{Galaxies:active--galaxies:jets--galaxies:general}
}

   \authorrunning{Yongyun Chen et al.}            
   \titlerunning{jets of supermassive black hole}  
   \maketitle

\section{INTRODUCTION}
The origin of jets from active galactic nuclei (AGN) has always been unclear. It is generally believed that the origin of jets is mainly related to accretion and the spin of supermassive black hole \citep[e.g.,]{blandford77, Ghisellini2006, Zamaninasab2014}. Some authors have suggested that the power of jets is closely linked to the accretion disk luminosity \citep[e.g.,][]{Rawlings91, Ghisellini14, Sbarrato14, Chen15, Che15, Paliya2017, Chen2023a, Chen2023b, Chen2023c}. \cite{Ghisellini14} found that the power of jets is greater than the luminosity of the accretion disk, which implies that the spin of black holes may play an important role in the formation of jets besides accretion. Some authors have found that the jets of stellar mass black holes in X-ray binaries are related to the black hole spin \citep{Narayan12, Steiner2013}. Recently, \cite{Cui2023} found a precessing jet nozzle connecting to a spinning black hole in M87. This result may suggest that the jet is closely related to the spin of the black hole. However, \cite{Fender10} found that there was no correlation between the spin of black holes and jets in X-ray binaries. Although there are some studies on the relationship between black hole spin and jet. However, whether the black hole spin enhances the relativistic jets has not been studied by using large samples with reliable black hole spins.

The origin of the dichotomy of radio-quiet and radio-loud AGN has been discussed in the literature. Some authors suggest that the accretion rate (Eddington ratios $\lambda=L_{\rm bol}/L_{\rm Edd}$, where $L_{\rm bol}$ is the bolometric luminosity and $L_{\rm Edd}$ is the Eddington luminosity) may explain this bimodal phenomenon \citep[e.g.,][]{Ho2000, Sikora2007}. Another possible explanation is that the black hole spin leads to the dichotomy phenomenon of radio-loud and radio-quiet AGN. Recently, \cite{Sikora2007} discovered that radio-loud AGN is hosted in elliptical galaxies, while radio-quiet AGN is likely hosted in spiral galaxies. \cite{Volonteri2007} found that the spiral galaxies have lower spin of black hole than elliptical galaxies. This result indicates that the dichotomy of radio-loud and radio-quiet AGN can be explained by the supermassive black hole spin. However, there is currently a lack of observational evidence to support the correlation between black hole spin and radio loudness.

In this paper, we mainly study the relationship between jet and black hole spin and accretion by using large samples. We also study the relationship between black hole spin and radio loudness. The sample is presented in Section 2; the jet model is described in Section 3; Section 4 presents the results and discussion; Section 5 describes the conclusions. A $\Lambda$CDM cosmology with $H_{0}=70 \rm km~s^{-1}Mpc^{-1}$, $\Omega_{\rm m}=0.27$, $\Omega_{\Lambda}=0.73$ is adopted.  

\section{The sample}
\subsection{The sample of Supermassive black hole}    
We select a large sample of supermassive black hole with reliably black hole spin, black hole mass, broad line region luminosity (BLR), and 1.4 GHz radio flux. The method of X-Ray Reflection Spectroscopy is used to measure the spin of a black hole \citep{Gallo2011, Walton2013, Risaliti2013, Lohfink2013, Brenneman2013, Reis2014, Keck2015, Parker2018, Sun2018, Buisson2018, Walton2019, Jiang2019, Walton2020}. Meanwhile, we consider that these sources have the luminosity of broad emission lines \citep{Sulentic2007, Sugai2007, Koss2017, Malkan2017, Grupe2004, Assef2011, Buttiglione2010, Daniel2018, Afanasiev2019, Kim2021}. The accretion disk luminosity is estimated by using ${L_{\rm disk}=10L_{\rm BLR}}$, with an average uncertainty of a factor 2 \citep{Calderone2013, Ghisellini14}. We also calculate the radio-loudness using the ratio of the flux of 1.4 GHz to the flux of B-band, $R=S_{1.4}/S_{B}$ \citep[e.g.,][]{Hao2014}. 

\subsection{The jet power}
The jet power of the AGN can be derived from the
radio luminosity of the compact core \citep[e.g.,][]{Merloni2007, Cavagnolo20, Daly2012}. \cite{Cavagnolo20} obtained the relationship between radio luminosity and jet power using 21 giant elliptical galaxies

\begin{equation}
	\log P_{\rm jet} = 0.75(\pm0.14)\log P_{1.4} + 1.91(\pm0.18),
\end{equation} 
where $L_{1.4}$ is the 1.4 GHz radio powers, which is estimated by the relation $P_{1.4}=4\pi d_{\rm L}^{2}(1+z)^{\alpha-1}S_{\nu}\nu$. The $S_{\nu}$ is the flux density at 1.4 GHz from NASA/IPAC Extragalactic Database (NED), and the spectral index $\alpha=0$ \citep[e.g.,][]{Abdo10, Komossa18}. The $P_{\rm jet}$ is in units $10^{42}$ erg s$^{-1}$ and $P_{1.4}$ in units $10^{40}$ erg s$^{-1}$. The scatter for this relation is $\sigma_{1.4} = 0.78$ dex. In the sample of \cite{Cavagnolo20}, about 76\% are radio-quiet AGN (e.g., NGC 4636, NGC 5813, and NGC 5846). Therefore, equation (1) can be used to calculate the jet power of radio-quiet AGN. Some authors also used the above equation to estimate the jet power of radio-quiet AGN \citep[e..g,][]{Cheung2016, Mezcua2019, Chen2020, Singha2023, Igo2024}. We also use equation (1) to estimate the jet power of our sample. We also note that the radio-quiet AGNs only show compact radio core, and radio flux mainly comes from the radio core. According to the radio luminosity of 1.4 GHz \citep[e.g.,][]{Fanaroff1974}, we find that almost all of our samples are FR I radio sources (L$_{\rm 1.4GHz}\lesssim10^{40.5}\rm erg~s^{-1}$). This suggests that our sample has bright core, and the 1.4 GHz radio flux mainly comes from core. Therefore, using the radio flux of 1.4 GHz to estimate jet power will not affect our main results. In the future, when these sources have a flux of radio core, we are testing our results.

\section{The jet model}
At present, the formation mechanism of jet mainly includes the Blandford-Znajek (BZ) mechanism \citep{blandford77}, Blandford Payne (BP) mechanism \citep{Blandford1982} and hybrid jet model \citep[e.g.,][]{Meier2001, Garofalo2010}, that is, the combination of BZ and BP.  It is generally believed that jets and/or outflows can be accelerated and collimated by large-scale magnetic fields \citep[e.g.,][]{Pudritz2007}. \cite{Spruit2005} suggested that the large-scale magnetic field that accelerates the jet or outflow is formed by the advection of a weak external field. However, \cite{Lubow1994} found that the advection of the external field caused by the small radial velocity of geometrically thin accretion disks (H/R$\ll$1) is quite ineffective. The coronal mechanism is proposed to alleviate the problem of advection of a weak external field in thin disks \citep[e.g.,][]{Spruit2005, Cao2013}. \cite{Beckwith2009} found that the hot corona above the accretion disk can effectively drag the external magnetic field to move inward. Some authors found that the magnetic field of the corona can enhance the relativistic jet \citep[e.g.,][]{Cao2018}. Our sample has high relativistic jet power, $\langle \log P_{\rm jet} \rangle =43.50\pm0.35$. In this work, we, therefore, use the magnetic field of the corona above the thin accretion disk to estimate the jet power of the model. Then, the jet power of the model is compared with the observed jet power.

\subsection{Magnetic field of  Corona}
The corona above the accretion disk can be described by using the the corona thickness ${H}_{\rm c}$ and the optical depth $\tau_{\rm c}$ \citep{Cao2018}. \cite{Cao2018} estimated the magnetic field strength of the corona using the following formula

\begin{equation}  
	B=4.37\times10^{8}\beta^{1/2}\tau_{\rm c}^{1/2}\tilde{H}_{\rm c}^{1/2}m^{-1/2}r^{-3/2}L_{*}^{2}~Gauss
\end{equation}

\begin{eqnarray}
	r = \frac{Rc^{2}}{GM_{\rm bh}},  m = \frac{M_{\rm bh}}{M_{\odot}}, \tilde{H}_{\rm c} = \frac{H_{\rm c}}{R}	
\end{eqnarray}
where $\tilde{H_{\rm c}}$ is the relative thickness \citep{Cao2018}. To estimate the maximum power of the jet, $\omega_{\rm F}=1/2$ \citep{Ghosh1997}, $\beta=1$, $\xi_{\phi}=1$, $\tilde{\Omega}=1$, $\tau_{\rm c}=0.5$, $\tilde{H_{\rm c}}=0.5$, and the $L_{*}=L_{\rm K}(r_{\rm ms})$ are used \citep{Cao2018}.

\subsection{The jet model}
(1) The BZ jet model.

The formula for calculating the jet power of the BZ mechanism is as follows \citep[e.g.,][]{MacDonald1982, Thorne1986, Ghosh1997, Livio1999, Nemmen2007}:

\begin{equation}
	P_{\rm jet}^{\rm BZ} = \frac{1}{32}\omega_{\rm F}^{2}B_{\perp}^{2}R_{\rm H}^{2}j^{2}c,
\end{equation}
where $B_{\perp}$ is the magnetic field strength of the black hole horizon, $B_{\perp}\approx B$, and $R_{\rm H}=[1+(1-j^{2})^{1/2}]GM_{\rm bh}/c^{2}$ indicates the horizon radius. The $\omega_{\rm F}$ is estimated by using the angular velocity of field lines $\Omega_{F}$ and the hole $\Omega_{\rm H}$, $\omega_{\rm F}\equiv \Omega_{\rm F}(\Omega_{\rm H}-\Omega_{\rm F})/\Omega_{\rm H}^{2}$. In order to obtain the maximal jet power of BZ mechanism, the $\omega_{\rm F}=1/2$ is used \citep{MacDonald1982}.

(2) The BP jet model.

We use the following formula to calculate the jet power of the BP jet model \citep{Livio1999, Cao2018}

\begin{equation}
	P_{\rm jet}^{\rm BP}\sim \frac{BB_{\rm \phi}^{s}}{2\pi}R_{\rm j}\Omega\pi R_{\rm j}^{2},
\end{equation}
where $B_{\rm \phi}^{\rm s}$ is $B_{\rm \phi}^{\rm s}=\xi_{\phi}B$. The $B_{\rm \phi}^{\rm s}$ indicates the azimuthal component of the magnetic field on the corona surface. The ratio $\xi_{\phi} \leq 1$ is required \citep{Livio1999}. The $R_{\rm j}$ indicates the radius of the jet formation area in the corona. The $\Omega$ is the angular velocity of the gas in the corona. According equation (2) and (5), the jet power of BP mechanism can be estimated as follows

\begin{equation}
	P_{\rm jet}^{BP}\simeq 3.13\times10^{37}\xi_{\phi}\tilde{\Omega}r_{\rm j}^{-1/2}m\beta\tau_{\rm c}\tilde{H_{\rm c}}~{\rm erg~s}^{-1}.	
\end{equation}
Due to the fact that most of the released gravitational power is located in the internal region of the accretion disk with a radius $\sim2R_{\rm ms}$ \citep{Shakura1973}, we use $R_{\rm j}=2R_{\rm ms}$ to estimate jet power of of BP model. The $R_{\rm ms}$ can be derived by the following formula,

\begin{eqnarray}
	R_{\rm ms} = R_{\rm G}\{3+Z_{2}-\left[(3-Z_{1})(3+Z_{1}+2Z_{2})\right]^{1/2}\},\nonumber \\
	Z_{1}\equiv1+(1-a^{2})^{1/3}\left[(1+a)^{1/3}+(1-a)^{1/3}\right], \nonumber \\
	Z_{2}\equiv(3a^{2}+Z_{1}^{2})^{1/2},\nonumber \\
	R_{\rm G} =\frac{GM_{\rm bh}}{c^{2}}.
\end{eqnarray}

(3) The Hybrid jet model. 

The hybrid model is a mixture of BZ and BP mechanisms. \cite{Garofalo2010} used the following formula to obtain the jet power of the hybrid model in the case of a thin accretion disk

\begin{equation}
	P_{\rm jet}^{\rm Hybrid} = 2\times10^{47} \alpha f^{2}\left(\frac{B_{\rm pd}}{10^{5}G}\right)^{2}\left(\frac{m}{10^{9}M_{\odot}}\right)^{2}j^{2} {\rm erg~s}^{-1},
\end{equation} 
where $B_{\rm pd}$ is $B_{\rm pd}\simeq B$. The BZ and BP mechanisms are combined with parameters $\alpha$ and $f$. The $\alpha$ and $f$ is respectively given by \citep{Garofalo2009}  
\begin{eqnarray}
	\alpha=\delta\left(\frac{3}{2}-j\right)	
\end{eqnarray}
and 
\begin{equation}
	\begin{split}
		f=-\frac{3}{2}j^{3}+12j^{2}-10j+7-\frac{0.002}{(j-0.65)^{2}}+\frac{0.1}{(j+0.95)}\\+\frac{0.002}{(j-0.055)^{2}}.
	\end{split}
\end{equation}
The conservative value of $\delta$ is about 2.5 \citep{Garofalo2010}. The $\alpha$ represents that the effectiveness of the BP jet is a function of the spin of black hole, while $f$ reflects the enhancing effect of the disk thread field on the black hole.

\section{Result and Discussion}
\subsection{The formation mechanism of jets}

\begin{figure*}
		\includegraphics[width=16.0cm,height=8.0cm]{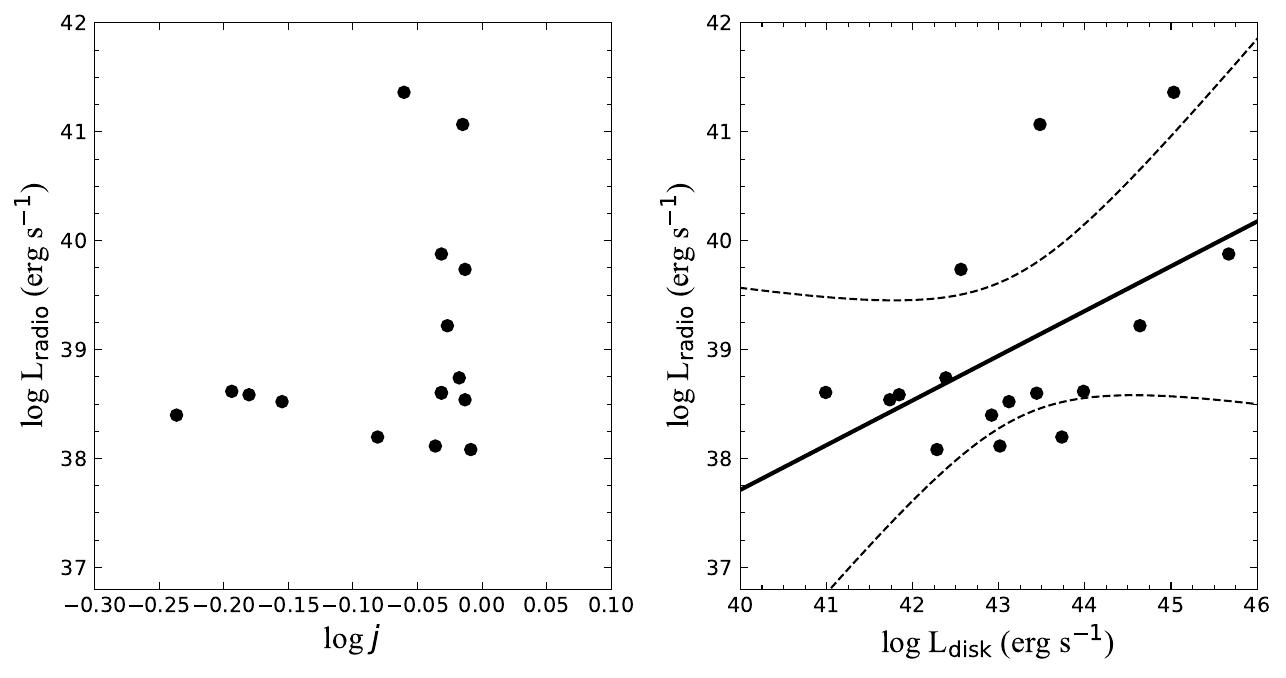}
	\centering
	\caption{Relation between black hole spin (left panel) and accretion disk luminosity (right panel) and radio luminosity for supermassive black hole. Solid line is the best linear fitting. Dashed lines is the 3$\sigma$ confidence band.
	}
	\label{pjetspin}
\end{figure*}

At present, there are three popular jet formation mechanisms, including BZ mechanism \citep{blandford77}, BP mechanism \citep{Blandford1982} and hybrid model \citep{Meier2001,Garofalo2010}. The BZ mechanism mainly extracts the rotational energy of the black hole, while the BP mechanism mainly extracts the rotational energy of the accretion disk. The hybrid model is a mixture of the BZ and BP mechanisms. In the BZ and BP mechanisms, the accretion of matter in the black hole leads to an expected relationship between the jet power and the accretion disk \citep{Maraschi2003}.

Figure.\ref{pjetspin} shows the relationship between black hole spin (left panel) and accretion disk luminosity (right panel) and radio luminosity for selected sample. From the left panel of Figure.\ref{pjetspin}, we find that there is a weak correlation between radio luminosity and black hole spin for selected sample (r=0.27). We also further test the correlation between the jet power and the black hole spin at an 84\% confidence level and find that the correlation coefficient was 0.42. This result shows a moderately strong correlation between jet power and black hole spin. The best-fit equation between radio luminosity and spin is $\log L_{\rm radio}=(3.57\pm3.39)\log j+(39.26\pm0.35)$ for selected sample. The result suggests that jet power depends on the spin of the black hole.

Some authors have found a close connection between jet and accretion \citep[e.g.,][]{Rawlings91, Ghisellini14, Sbarrato14, Chen15, Che15, Paliya2017, Mukherjee19, Chen2023a, Chen2023b, Chen2023c}. We test this correlation. The right panel of Figure.\ref{pjetspin} shows a relation between radio luminosity and accretion disk luminosity for selected sample. We find a significant correlation between the luminosity of radio and the luminosity of accretion disk for selected sample ($r=0.51, p=0.04$). This result further proves that there is a close connection between jet and accretion for our sample. The best-fit equation between radio luminosity and accretion disk luminosity is $\log L_{\rm radio}=(0.41\pm0.18)\log L_{\rm disk}+(21.28\pm7.91)$ for selected sample. The results in Figure~\ref{pjetspin} may further imply that the spin and accretion of the black hole enhance the relativistic jet. The jet model of our sample may be a hybrid model. We test this speculation.

\begin{figure}
	\includegraphics[width=0.85\columnwidth]{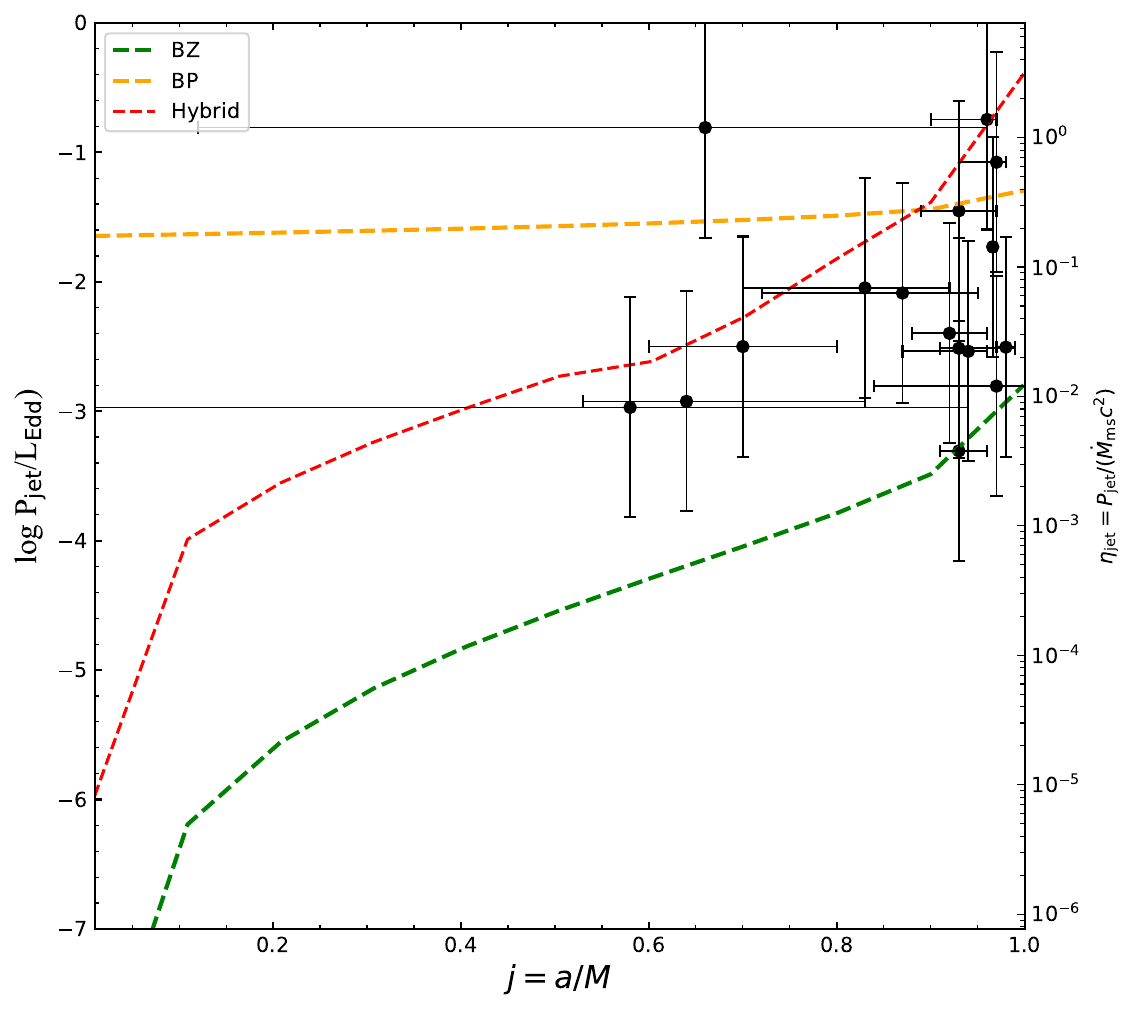}
	\centering
	\caption{
		The jet power in Eddington units as a function of black hole spin. 
		The orange dashed line indicates the jet power of BP mechanism. The green dashed line indicates the BZ mechanism. The red dashed line indicates the hybrid model. The right-hand axis indicates the jet efficiency. The uncertainty of $\log P_{\rm jet}/L_{\rm Edd}$ mainly comes from $\log P_{\rm jet}$.
	}
	\label{jetmodel}
\end{figure}

The relationship between jet power in Eddington units of the model and black hole spin is shown in Figure.\ref{jetmodel}. The green dashed line is BZ mechanism. The red dashed line is Hybrid model. The orange line is BP mechanism. From Figure.\ref{jetmodel}, we find that the jet power of the BZ mechanism and the hybrid model varies with the spin of the black hole, while the jet power of the BP mechanism does not change with the spin of the black hole. The greater the spin of a black hole, the higher the efficiency of the jet, which is consistent with GRMHD simulations \citep{Tchekhovskoy12}. The jet efficiency of BZ model and the hybrid model is 1\% and 300\% when the black hole spin is 0.99, respectively. This result imply that the hybrid model plays an important role than the BZ mechanism. At the same time, we find that the jet power of most selected sample is below the red dashed line and orange dashed line except Mrk 359, which indicates that the hybrid model can explain the jet power of almost all selected sample. We also note that Mrk359 has a relatively low black hole mass, resulting in a high $\log P_{\rm jet}/L_{\rm Edd}$. 

\begin{figure}
	\includegraphics[width=0.85\columnwidth]{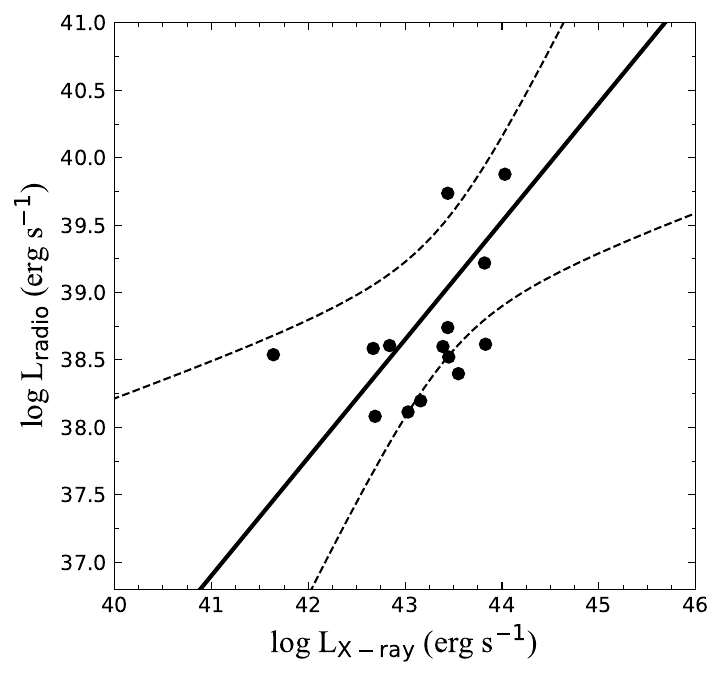}
	\centering
	\caption{The radio luminosity as a function of hard X-ray luminosity for supermassive black hole. The physical meaning of black solid and dashed lines is the same as in figure 1.
	}
	\label{LradioLX}
\end{figure}

The properties of a coronae are usually closely related to the emission of hard X-rays.
If the jet model of the corona does indeed work in these supermassive black holes, one possibility is to expect a correlation between radio and hard X-ray luminosity. We check this correlation. The relation between radio and hard X-ray luminosity is shown in Figure.\ref{LradioLX}. We find a significant correlation between radio luminosity and hard X-ray luminosity ($r=0.75, p=0.0009$). 

\begin{figure}
	\includegraphics[width=0.85\columnwidth]{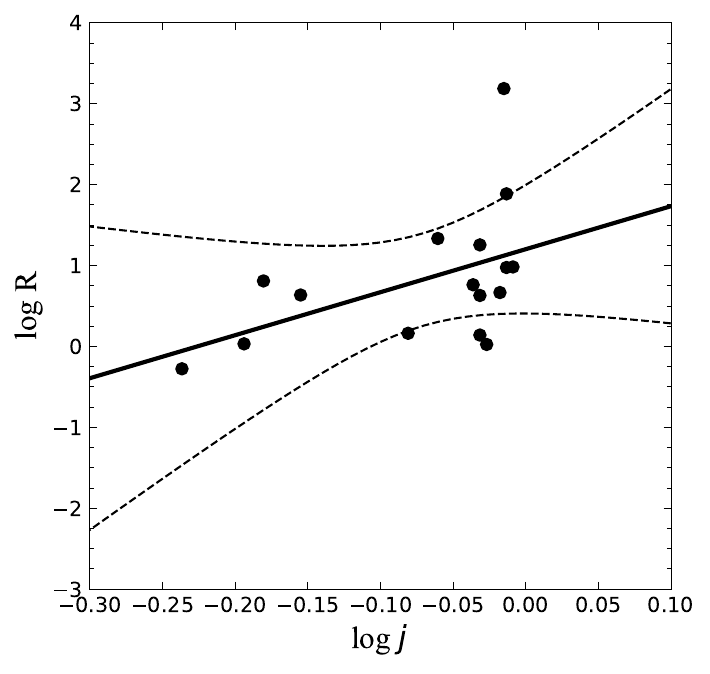}
	\centering
	\caption{Relation radio loudness and  black hole spin for selected sample.
		The physical meaning of black solid and dashed lines is the same as in figure 1.}
	\label{Rj}
\end{figure}

\subsection{Relation between black hole spin and radio loudness}
AGN is divided into two types: radio-loud and radio-quiet AGN based on radio loudness. The origin of radio loudness has been unclear. Some studies have found that there is an inverse correlation between radio-loudness and accretion \citep[e.g.,][]{Sikora2007}. In addition to the accretion rate, it is natural that the second physical parameter that determines the radio-loudness of AGN is related to the properties of the central black hole, such as the spin of a black hole. We test this correlation. Figure~\ref{Rj} shows the relation between radio-loudness and black hole spin for supermassive black holes. We find a significant between radio-loudness and spin of black holes for selected sample ($r=0.48, p=0.04$). The best-fit equation between radio radio-loudness and spin is $\log R=(5.29\pm2.60)\log j+(1.18\pm0.26)$ for selected sample. The sources with high radio-loudness tend to have high black hole spin, which may imply that the spin of black holes can explain the dichotomy of radio-quiet and radio-loud AGN. \cite{Tchekhovskoy2010} investigate whether the bimodal of radio-loud and radio-quiet can be due to differences in the black hole spin by using numerical simulation. They suggested that the dichotomy of radio-loud and radio-quiet can be explained by two different populations of galaxies with modestly different black hole spins. The radio-loud and radio-quiet AGN may have different merger and accretion histories, which lead to different black hole spin. The radio-loud AGN hosts in elliptical galaxies, whereas radio-quiet AGN is likely to host in spiral galaxies \citep{Sikora2007}. \cite{Volonteri2007} found that the average spin of supermassive black holes in elliptical galaxies is higher than that in spiral galaxies. The radio-loud AGN may have higher black hole spin than radio-quiet AGN. However, we find the parameter space for the black hole spin is quite narrow (e.g., log$j$  from -0.3 to 0), where most sources have very high spin parameter and only 4 sources with j$\sim$0.6-0.7. This correlation may be strongly affected by the sample selection. In the future, our results should be tested using large samples.

\section{Conclusions}
In this work, we study the relation between jet, accretion, and black hole spin. We also study the relation between radio loudness and black hole spin. Our main results are as follows:

1. We find that there is a weak correlation between radio luminosity and the black hole spin for our sample. We also further test the correlation between the jet power and the black hole spin at an 84\% confidence level and find that the correlation coefficient was 0.42. These results imply that the jet of the supermassive black hole in our sample mainly depends on the other parameters besides black hole spin.

2. We investigate the jet power of the BZ mechanism, BP mechanism, and 
hybrid mechanism based on the thin disk surrounding Kerr black holes. According to the diagram of the relationship between jet power in Eddington units and black hole spin, we find that the hybrid model can explain the jet power of almost all sources.

3. There is a significant correlation between radio loudness and black hole spin for our sample. This result may provide observational evidence for explaining radio-quiet and radio-loud dichotomy. 

\begin{acknowledgements}
We are very grateful to referee and Editor for the very helpful report. Yongyun Chen is grateful for financial support from the National Natural Science Foundation of China (No. 12203028). This work was support from the research project of Qujing Normal University (2105098001/094). This work is supported by the youth project of Yunnan Provincial Science and Technology Department (202101AU070146, 2103010006). Yongyun Chen is grateful for funding for the training Program for talents in Xingdian, Yunnan Province. 
QSGU is supported by the National Natural Science Foundation of China (12121003, 12192220, and 12192222).
We also acknowledge the science research grants from the China Manned Space Project with NO. CMS-CSST-2021-A05. This work is supported by the National Natural Science Foundation of China (11733001, U2031201 and 12433004). 
\end{acknowledgements}

\bibliographystyle{raa}
\bibliography{example}

\begin{thebibliography}{88}
\providecommand\natexlab[1]{#1}
\providecommand\JournalTitle[1]{#1}

\bibitem[{Abdo} {et~al.}(2010)]{Abdo10}
{Abdo}, A.~A., {Ackermann}, M., {Agudo}, I., {et~al.} 2010, \apj, 716, 30

\bibitem[{Afanasiev} {et~al.}(2019)]{Afanasiev2019}
{Afanasiev}, V.~L., {Popovi{\'c}}, L.~{\v{C}}., \& {Shapovalova}, A.~I. 2019,
  \mnras, 482, 4985

\bibitem[{Assef} {et~al.}(2011)]{Assef2011}
{Assef}, R.~J., {Denney}, K.~D., {Kochanek}, C.~S., {et~al.} 2011, \apj, 742,
  93

\bibitem[{Beckwith} {et~al.}(2009)]{Beckwith2009}
{Beckwith}, K., {Hawley}, J.~F., \& {Krolik}, J.~H. 2009, \apj, 707, 428

\bibitem[{Bennert} {et~al.}(2011)]{Bennert2011}
{Bennert}, V.~N., {Auger}, M.~W., {Treu}, T., {Woo}, J.-H., \& {Malkan}, M.~A.
  2011, \apj, 726, 59

\bibitem[{Blandford} \& {Payne}(1982)]{Blandford1982}
{Blandford}, R.~D., \& {Payne}, D.~G. 1982, \mnras, 199, 883

\bibitem[{Blandford} \& {Znajek}(1977)]{blandford77}
{Blandford}, R.~D., \& {Znajek}, R.~L. 1977, \mnras, 179, 433

\bibitem[{Brenneman}(2013)]{Brenneman2013}
{Brenneman}, L. 2013, {Measuring the Angular Momentum of Supermassive Black
  Holes}

\bibitem[{Buisson} {et~al.}(2018)]{Buisson2018}
{Buisson}, D.~J.~K., {Parker}, M.~L., {Kara}, E., {et~al.} 2018, \mnras, 480,
  3689

\bibitem[{Buttiglione} {et~al.}(2010)]{Buttiglione2010}
{Buttiglione}, S., {Capetti}, A., {Celotti}, A., {et~al.} 2010, \aap, 509, A6

\bibitem[{Calderone} {et~al.}(2013)]{Calderone2013}
{Calderone}, G., {Ghisellini}, G., {Colpi}, M., \& {Dotti}, M. 2013, \mnras,
  431, 210

\bibitem[{Cao}(2018)]{Cao2018}
{Cao}, X. 2018, \mnras, 473, 4268

\bibitem[{Cao} \& {Spruit}(2013)]{Cao2013}
{Cao}, X., \& {Spruit}, H.~C. 2013, \apj, 765, 149

\bibitem[{Cavagnolo} {et~al.}(2010)]{Cavagnolo20}
{Cavagnolo}, K.~W., {McNamara}, B.~R., {Nulsen}, P.~E.~J., {et~al.} 2010, \apj,
  720, 1066

\bibitem[{Chen} {et~al.}(2020)]{Chen2020}
{Chen}, S., {J{\"a}rvel{\"a}}, E., {Crepaldi}, L., {et~al.} 2020, \mnras, 498,
  1278

\bibitem[{Chen} {et~al.}(2023{\natexlab{a}})]{Chen2023c}
{Chen}, Y., {Gu}, Q., {Fan}, J., {et~al.} 2023{\natexlab{a}}, \mnras, 519, 6199

\bibitem[{Chen} {et~al.}(2023{\natexlab{b}})]{Chen2023a}
{Chen}, Y., {Gu}, Q., {Fan}, J., {et~al.} 2023{\natexlab{b}}, \apjs, 265, 60

\bibitem[{Chen} {et~al.}(2015{\natexlab{a}})]{Che15}
{Chen}, Y.-Y., {Zhang}, X., {Xiong}, D., \& {Yu}, X. 2015{\natexlab{a}}, \aj,
  150, 8

\bibitem[{Chen} {et~al.}(2015{\natexlab{b}})]{Chen15}
{Chen}, Y.~Y., {Zhang}, X., {Zhang}, H.~J., \& {Yu}, X.~L. 2015{\natexlab{b}},
  \mnras, 451, 4193

\bibitem[{Chen} {et~al.}(2023{\natexlab{c}})]{Chen2023b}
{Chen}, Y., {Gu}, Q., {Fan}, J., {et~al.} 2023{\natexlab{c}}, \apjs, 268, 6

\bibitem[{Cheung} {et~al.}(2016)]{Cheung2016}
{Cheung}, E., {Bundy}, K., {Cappellari}, M., {et~al.} 2016, \nat, 533, 504

\bibitem[{Cui} {et~al.}(2023)]{Cui2023}
{Cui}, Y., {Hada}, K., {Kawashima}, T., {et~al.} 2023, \nat, 621, 711

\bibitem[{Daly} {et~al.}(2012)]{Daly2012}
{Daly}, R.~A., {Sprinkle}, T.~B., {O'Dea}, C.~P., {Kharb}, P., \& {Baum}, S.~A.
  2012, \mnras, 423, 2498

\bibitem[{Daniel} \& {Wyse}(2018)]{Daniel2018}
{Daniel}, K.~J., \& {Wyse}, R. F.~G. 2018, \mnras, 476, 1561

\bibitem[{Event Horizon Telescope Collaboration} {et~al.}(2019)]{Akiyama2019}
{Event Horizon Telescope Collaboration}, {Akiyama}, K., {Alberdi}, A., {et~al.}
  2019, \apjl, 875, L1

\bibitem[{Fanaroff} \& {Riley}(1974)]{Fanaroff1974}
{Fanaroff}, B.~L., \& {Riley}, J.~M. 1974, \mnras, 167, 31P

\bibitem[{Fender} {et~al.}(2010)]{Fender10}
{Fender}, R.~P., {Gallo}, E., \& {Russell}, D. 2010, \mnras, 406, 1425

\bibitem[{Gallo} {et~al.}(2011)]{Gallo2011}
{Gallo}, L.~C., {Miniutti}, G., {Miller}, J.~M., {et~al.} 2011, \mnras, 411,
  607

\bibitem[{Garofalo}(2009)]{Garofalo2009}
{Garofalo}, D. 2009, \apj, 699, 400

\bibitem[{Garofalo} {et~al.}(2010)]{Garofalo2010}
{Garofalo}, D., {Evans}, D.~A., \& {Sambruna}, R.~M. 2010, \mnras, 406, 975

\bibitem[{Ghisellini}(2006)]{Ghisellini2006}
{Ghisellini}, G. 2006, in VI Microquasar Workshop: Microquasars and Beyond,
  27.1

\bibitem[{Ghisellini} {et~al.}(2014)]{Ghisellini14}
{Ghisellini}, G., {Tavecchio}, F., {Maraschi}, L., {Celotti}, A., \&
  {Sbarrato}, T. 2014, \nat, 515, 376

\bibitem[{Ghosh} \& {Abramowicz}(1997)]{Ghosh1997}
{Ghosh}, P., \& {Abramowicz}, M.~A. 1997, \mnras, 292, 887

\bibitem[{Grier} {et~al.}(2017)]{Grier2017}
{Grier}, C.~J., {Pancoast}, A., {Barth}, A.~J., {et~al.} 2017, \apj, 849, 146

\bibitem[{Grupe} {et~al.}(2004)]{Grupe2004}
{Grupe}, D., {Mathur}, S., {Wilkes}, B., \& {Elvis}, M. 2004, \aj, 127, 1

\bibitem[{Hao} {et~al.}(2014)]{Hao2014}
{Hao}, H., {Sargent}, M.~T., {Elvis}, M., {et~al.} 2014, arXiv e-prints,
  arXiv:1408.1090

\bibitem[{Ho} {et~al.}(2000)]{Ho2000}
{Ho}, L.~C., {Rudnick}, G., {Rix}, H.-W., {et~al.} 2000, \apj, 541, 120

\bibitem[{Igo} {et~al.}(2024)]{Igo2024}
{Igo}, Z., {Merloni}, A., {Hoang}, D., {et~al.} 2024, \aap, 686, A43

\bibitem[{Jiang} {et~al.}(2019)]{Jiang2019}
{Jiang}, J., {Walton}, D.~J., {Fabian}, A.~C., \& {Parker}, M.~L. 2019, \mnras,
  483, 2958

\bibitem[{Keck} {et~al.}(2015)]{Keck2015}
{Keck}, M.~L., {Brenneman}, L.~W., {Ballantyne}, D.~R., {et~al.} 2015, \apj,
  806, 149

\bibitem[{Kim} {et~al.}(2021)]{Kim2021}
{Kim}, M., {Barth}, A.~J., {Ho}, L.~C., \& {Son}, S. 2021, \apjs, 256, 40

\bibitem[{Komossa} {et~al.}(2018)]{Komossa18}
{Komossa}, S., {Xu}, D.~W., \& {Wagner}, A.~Y. 2018, \mnras, 477, 5115

\bibitem[{Koss} {et~al.}(2017)]{Koss2017}
{Koss}, M., {Trakhtenbrot}, B., {Ricci}, C., {et~al.} 2017, \apj, 850, 74

\bibitem[{Lee} {et~al.}(2013)]{Lee2013}
{Lee}, N., {Sanders}, D.~B., {Casey}, C.~M., {et~al.} 2013, \apj, 778, 131

\bibitem[{Livio} {et~al.}(1999)]{Livio1999}
{Livio}, M., {Ogilvie}, G.~I., \& {Pringle}, J.~E. 1999, \apj, 512, 100

\bibitem[{Lohfink} {et~al.}(2013)]{Lohfink2013}
{Lohfink}, A.~M., {Reynolds}, C.~S., {Jorstad}, S.~G., {et~al.} 2013, \apj,
  772, 83

\bibitem[{Lubow} {et~al.}(1994)]{Lubow1994}
{Lubow}, S.~H., {Papaloizou}, J.~C.~B., \& {Pringle}, J.~E. 1994, \mnras, 267,
  235

\bibitem[{MacDonald} \& {Thorne}(1982)]{MacDonald1982}
{MacDonald}, D., \& {Thorne}, K.~S. 1982, \mnras, 198, 345

\bibitem[{Malkan} {et~al.}(2017)]{Malkan2017}
{Malkan}, M.~A., {Jensen}, L.~D., {Rodriguez}, D.~R., {Spinoglio}, L., \&
  {Rush}, B. 2017, \apj, 846, 102

\bibitem[{Maraschi} \& {Tavecchio}(2003)]{Maraschi2003}
{Maraschi}, L., \& {Tavecchio}, F. 2003, \apj, 593, 667

\bibitem[{Meier}(2001)]{Meier2001}
{Meier}, D.~L. 2001, \apjl, 548, L9

\bibitem[{Mel{\'e}ndez} {et~al.}(2010)]{Melendez2010}
{Mel{\'e}ndez}, M., {Kraemer}, S.~B., \& {Schmitt}, H.~R. 2010, \mnras, 406,
  493

\bibitem[{Merloni} \& {Heinz}(2007)]{Merloni2007}
{Merloni}, A., \& {Heinz}, S. 2007, \mnras, 381, 589

\bibitem[{Mezcua} {et~al.}(2019)]{Mezcua2019}
{Mezcua}, M., {Suh}, H., \& {Civano}, F. 2019, \mnras, 488, 685

\bibitem[{Mukherjee} {et~al.}(2019)]{Mukherjee19}
{Mukherjee}, S., {Mitra}, K., \& {Chatterjee}, R. 2019, \mnras, 486, 1672

\bibitem[{Narayan} \& {McClintock}(2012)]{Narayan12}
{Narayan}, R., \& {McClintock}, J.~E. 2012, \mnras, 419, L69

\bibitem[{Nemmen} {et~al.}(2007)]{Nemmen2007}
{Nemmen}, R.~S., {Bower}, R.~G., {Babul}, A., \& {Storchi-Bergmann}, T. 2007,
  \mnras, 377, 1652

\bibitem[{Niko{\l}ajuk} {et~al.}(2009)]{Nikolajuk2009}
{Niko{\l}ajuk}, M., {Czerny}, B., \& {Gurynowicz}, P. 2009, \mnras, 394, 2141

\bibitem[{Onken} {et~al.}(2014)]{Onken2014}
{Onken}, C.~A., {Valluri}, M., {Brown}, J.~S., {et~al.} 2014, \apj, 791, 37

\bibitem[{Paliya} {et~al.}(2017)]{Paliya2017}
{Paliya}, V.~S., {Marcotulli}, L., {Ajello}, M., {et~al.} 2017, \apj, 851, 33

\bibitem[{Parker} {et~al.}(2018)]{Parker2018}
{Parker}, M.~L., {Matzeu}, G.~A., {Guainazzi}, M., {et~al.} 2018, \mnras, 480,
  2365

\bibitem[{Peterson} {et~al.}(2004)]{Peterson2004}
{Peterson}, B.~M., {Ferrarese}, L., {Gilbert}, K.~M., {et~al.} 2004, \apj, 613,
  682

\bibitem[{Pudritz} {et~al.}(2007)]{Pudritz2007}
{Pudritz}, R.~E., {Ouyed}, R., {Fendt}, C., \& {Brandenburg}, A. 2007, in
  Protostars and Planets V, ed. B.~{Reipurth}, D.~{Jewitt}, \& K.~{Keil}, 277

\bibitem[{Rawlings} \& {Saunders}(1991)]{Rawlings91}
{Rawlings}, S., \& {Saunders}, R. 1991, \nat, 349, 138

\bibitem[{Reis} {et~al.}(2014)]{Reis2014}
{Reis}, R.~C., {Reynolds}, M.~T., {Miller}, J.~M., \& {Walton}, D.~J. 2014,
  \nat, 507, 207

\bibitem[{Risaliti} {et~al.}(2013)]{Risaliti2013}
{Risaliti}, G., {Harrison}, F.~A., {Madsen}, K.~K., {et~al.} 2013, \nat, 494,
  449

\bibitem[{Sbarrato} {et~al.}(2014)]{Sbarrato14}
{Sbarrato}, T., {Padovani}, P., \& {Ghisellini}, G. 2014, \mnras, 445, 81

\bibitem[{Shakura} \& {Sunyaev}(1973)]{Shakura1973}
{Shakura}, N.~I., \& {Sunyaev}, R.~A. 1973, \aap, 24, 337

\bibitem[{Sikora} {et~al.}(2007)]{Sikora2007}
{Sikora}, M., {Stawarz}, {\L}., \& {Lasota}, J.-P. 2007, \apj, 658, 815

\bibitem[{Singha} {et~al.}(2023)]{Singha2023}
{Singha}, M., {Winkel}, N., {Vaddi}, S., {et~al.} 2023, \apj, 959, 107

\bibitem[{Spruit} \& {Uzdensky}(2005)]{Spruit2005}
{Spruit}, H.~C., \& {Uzdensky}, D.~A. 2005, \apj, 629, 960

\bibitem[{Steiner} {et~al.}(2013)]{Steiner2013}
{Steiner}, J.~F., {McClintock}, J.~E., \& {Narayan}, R. 2013, \apj, 762, 104

\bibitem[{Sugai} {et~al.}(2007)]{Sugai2007}
{Sugai}, H., {Kawai}, A., {Shimono}, A., {et~al.} 2007, \apj, 660, 1016

\bibitem[{Sulentic} {et~al.}(2007)]{Sulentic2007}
{Sulentic}, J.~W., {Bachev}, R., {Marziani}, P., {Negrete}, C.~A., \&
  {Dultzin}, D. 2007, \apj, 666, 757

\bibitem[{Sun} {et~al.}(2018)]{Sun2018}
{Sun}, S., {Guainazzi}, M., {Ni}, Q., {et~al.} 2018, \mnras, 478, 1900

\bibitem[{Tamburini} {et~al.}(2020)]{Tamburini2020}
{Tamburini}, F., {Thid{\'e}}, B., \& {Della Valle}, M. 2020, \mnras, 492, L22

\bibitem[{Tan} {et~al.}(2012)]{Tan2012}
{Tan}, Y., {Wang}, J.~X., {Shu}, X.~W., \& {Zhou}, Y. 2012, \apjl, 747, L11

\bibitem[{Tchekhovskoy} {et~al.}(2012)]{Tchekhovskoy12}
{Tchekhovskoy}, A., {McKinney}, J.~C., \& {Narayan}, R. 2012, in Journal of
  Physics Conference Series, Vol. 372, Journal of Physics Conference Series,
  012040

\bibitem[{Tchekhovskoy} {et~al.}(2010)]{Tchekhovskoy2010}
{Tchekhovskoy}, A., {Narayan}, R., \& {McKinney}, J.~C. 2010, \apj, 711, 50

\bibitem[{Thorne} {et~al.}(1986)]{Thorne1986}
{Thorne}, K.~S., {Price}, R.~H., \& {MacDonald}, D.~A. 1986, {Black holes: The
  membrane paradigm}

\bibitem[{{\"U}nal} \& {Loeb}(2020)]{Unal2020}
{{\"U}nal}, C., \& {Loeb}, A. 2020, \mnras, 495, 278

\bibitem[{Vasudevan} {et~al.}(2016)]{Vasudevan2016}
{Vasudevan}, R.~V., {Fabian}, A.~C., {Reynolds}, C.~S., {et~al.} 2016, \mnras,
  458, 2012

\bibitem[{Volonteri} {et~al.}(2007)]{Volonteri2007}
{Volonteri}, M., {Sikora}, M., \& {Lasota}, J.-P. 2007, \apj, 667, 704

\bibitem[{Walton} {et~al.}(2013)]{Walton2013}
{Walton}, D.~J., {Nardini}, E., {Fabian}, A.~C., {Gallo}, L.~C., \& {Reis},
  R.~C. 2013, \mnras, 428, 2901

\bibitem[{Walton} {et~al.}(2019)]{Walton2019}
{Walton}, D.~J., {Nardini}, E., {Gallo}, L.~C., {et~al.} 2019, \mnras, 484,
  2544

\bibitem[{Walton} {et~al.}(2020)]{Walton2020}
{Walton}, D.~J., {Alston}, W.~N., {Kosec}, P., {et~al.} 2020, \mnras, 499, 1480

\bibitem[{Zamaninasab} {et~al.}(2014)]{Zamaninasab2014}
{Zamaninasab}, M., {Clausen-Brown}, E., {Savolainen}, T., \& {Tchekhovskoy}, A.
  2014, \nat, 510, 126

\bibitem[{Zhou} \& {Wang}(2005)]{Zhou2005}
{Zhou}, X.-L., \& {Wang}, J.-M. 2005, \apjl, 618, L83

\end{thebibliography}

\begin{landscape}
	\begin{table}
		\caption{The sample of supermassive black hole}
		\centering
		\label{table1}
		\begin{tabular}{llllllllllllllllllllllll}
			\hline\hline
			Name & $z$ &  $S_{\nu}$  &$\log L_{\rm radio}$ & $\log M/M_{\odot}$ & $j$ &  Mass/spin Reference    &$\log L_{\rm BLR}$  & Ref & $S_{B}$ & $\log$ R & $S_{\rm X-ray}$ & $S_{\rm X-ray, error}$ \\
			~[1]  &[2]   &[3]   &[4]   &[5]   &[6]   &[7]   &[8]  &[9]   &[10]   &[11]  &[12] &[13]\\
           \hline
Mrk 335	&	0.02578	&	7.6$\pm$0.6	&	38.2	&	7.15$\pm$0.11	&	0.83$_{-0.13}^{+0.09}$	&	Pe04/Wa13	&	42.73	&	S07	&	5.22$\pm$0.173	&	0.16	&	6.50E-07	&	6.90E-09	\\
IRAS 00521-7054	&	0.069	&	36$\pm$2	&	39.74	&	7.7$\pm$0.09	&	0.98$_{0.04}^{+0.018}$	&	Wa19	&	41.56	&	M17	&	0.467$\pm$0.0495	&	1.89	&	1.66E-07	&	0.00E+00	\\
Mrk 359	&	0.0168	&	44$\pm$0.6	&	38.59	&	6.04	&	0.66$_{-0.54}^{+0.30}$	&	ZW05/Wa13	&	40.84	&	K17	&	6.81$\pm$0.727	&	0.81	&	5.10E-07	&	2.07E-08	\\
Mrk 1018	&	0.043	&	4.3$\pm$0.5	&	38.4	&	8.15	&	0.58$_{-0.74}^{+0.36}$	&	Be11/Wa13	&	41.92	&	B10	&	8.12$\pm$1.2	&	-0.28	&	5.59E-07	&	7.59E-08	\\
NGC 1365	&	0.00546	&	377$\pm$13	&	38.54	&	6.3$_{-0.23}^{+0.53}$	&	0.97$_{-0.04}^{+0.01}$	&	Fa19/Ri13	&	40.73	&	M17	&	39.8$\pm$0.79	&	0.98	&	4.55E-07	&	6.54E-08	\\
3C120	&	0.033	&	3440$\pm$103	&	41.07	&	7.74$_{-0.15}^{0.20}$	&	0.994$_{-0.04}^{+0.004}$	&	Pe04/Lo13	&	42.48	&	M17	&	2.23$\pm$0.41	&	3.19	&	3.05E-06	&	1.53E-07	\\
Ark120	&	0.0327	&	12.4$\pm$0.6	&	38.62	&	8.18$\pm$0.06	&	0.64$_{-0.11}^{+0.19}$	&	Pe04/Wa13	&	42.98	&	K17	&	11.5$\pm$0.46	&	0.03	&	1.89E-06	&	1.88E-07	\\
Mrk  79	&	0.022	&	22.2$\pm$1.3	&	38.52	&	7.72$\pm$0.12	&	0.7$_{-0.10}^{+0.10}$	&	Pe04/Ga11	&	42.12	&	M17	&	5.13$\pm$0.12	&	0.64	&	1.79E-06	&	1.79E-07	\\
NGC 3783	&	0.00973	&	44.6$\pm$2	&	38.11	&	7.47$\pm$0.08	&	0.92$_{-0.04}^{+0.04}$	&	Pe04/Br13	&	42.01	&	K17	&	7.7$\pm$0.24	&	0.76	&	3.55E-06	&	1.76E-07	\\
NGC 4151	&	0.0033	&	360$\pm$10.8	&	38.08	&	7.57$\pm$0.15	&	0.94$_{-0.05}^{+0.05}$	&	On14/Ke15	&	41.28	&	M17	&	37.4$\pm$0.61	&	0.98	&	1.39E-05	&	6.90E-07	\\
Mrk 766	&	0.01288	&	40.4$\pm$1.9	&	38.32	&	6.25$_{-0.04}^{+0.05}$	&	0.92$_{-0.05}^{+0.05}$	&	Be06/Bu18	&	40.36	&	K17	&	1.58$\pm$0	&	1.41	&	2.07E-06	&	0.00E+00	\\
PG1229+204	&	0.0636	&	3.1$\pm$0.5	&	38.6	&	7.76$\pm$0.22	&	0.93$_{-0.02}^{+0.06}$	&	Ji19	&	42.44	&	K17	&	2.24$\pm$1.31	&	0.14	&	1.75E-07	&	1.10E-08	\\
IRAS13349+2438	&	0.108	&	20$\pm$0.8	&	39.88	&	8.63$_{-0.17}^{+0.09}$	&	0.93$_{-0.02}^{+0.03}$	&	Pa18	&	44.67	&	A19	&	4.68$\pm$0	&	0.63	&	2.48E-07	&	0.00E+00	\\
Ark564	&	0.0247	&	29.1$\pm$1	&	38.74	&	6.04	&	0.96$_{-0.01}^{+0.11}$	&	ZW05/Wa13	&	41.39	&	S07	&	6.27$\pm$1.48	&	0.67	&	1.38E-06	&	0.00E+00	\\
NGC5506	&	0.00608	&	355$\pm$10	&	38.61	&	6.7$_{-0.10}^{+0.19}$	&	0.93$_{-0.04}^{+0.04}$	&	Ni09/Su18	&	40	&	K17	&	19.7$\pm$2.1	&	1.26	&	5.78E-06	&	0.00E+00	\\
IRAS09149-6206 	&	0.0573	&	16$pm$0	&	39.22	&	8$_{-0.3}^{+1.3}$	&	0.94$_{-0.07}^{+0.02}$	&	Wa20/Wa20	&	43.64	&	K17	&	15.1$\pm$0	&	0.03	&	5.86E-07	&	0.00E+00	\\
RXS J1131-1231 	&	0.654	&	15$\pm$0	&	41.36	&	8.3$\pm$0.09	&	0.87$_{-0.15}^{+0.08}$	&	Sl12/Re14	&	44.03	&	S07	&	0.694$\pm$0	&	1.33	&	1.66E-07	&	0.00E+00	\\          
			\hline
		\end{tabular}
	\end{table}
	\footnotesize{{\bf Notes.} Col. 1: name;
		Col. 2: redshift;
		Col. 3: the 1.4 GHz flux density in units mjy;
		Col. 4: Logarithm of 1.4 GHz radio luminosity (in units of erg s$^{-1}$);
		Col. 5: Logarithm of Black hole mass (in units of solar mass);
		Col. 6: The spin of black hole;
		Col. 7: the reference of spin and mass of black hole. Pe04= \cite{Peterson2004}; Wa13=\cite{Walton2013}; Ta12=\cite{Tan2012}; ZW05=\cite{Zhou2005}; Be11=\cite{Bennert2011}; Va16=\cite{Vasudevan2016}; Ri13=\cite{Risaliti2013}; Lo13=\cite{Lohfink2013}; Ga11=\cite{Gallo2011}; Un20=\cite{Unal2020}; On14=\cite{Onken2014}; Ke15=\cite{Keck2015}; Ta20=\cite{Tamburini2020}; Ak19=\cite{Akiyama2019}; Ja19=\cite{Jiang2019}; Pa18=\cite{Parker2018}; Ni09=\cite{Nikolajuk2009}; Su18=\cite{Sun2018}; Me10 = \cite{Melendez2010}; Gr17=\cite{Grier2017}
		Col. 8: Logarithm of broad region luminosity (in units of erg s$^{-1}$);
		Col. 9: the reference of broad line region luminosity. S07:  \cite{Sulentic2007}; M17=\cite{Malkan2017}; K17=\cite{Koss2017}; K18= \cite{Komossa18}; B10= \cite{Buttiglione2010}; A19= \cite{Afanasiev2019}; Le13=\cite{Lee2013}.
		Col. 10: the flux of optical B-band in units mjy comes from NED;
		Col .11: the radio-loudness;
		Col .12: the flux of X-ray 2-10 keV in units Jy comes from NED.
		Col .13: the flux error of X-ray 2-10 keV in units Jy comes from NED.}
\end{landscape}

\end{document}